\documentstyle[preprint,aps,12pt]{revtex}
\begin{document}
\title{
TESTING COMPLETE POSITIVITY}
\author{
Fabio Benatti
}
\address{
Dipartimento di Fisica Teorica, Universit\`a di Trieste\\
Strada Costiera 11, 34014 Trieste, Italy\\
and\\
Istituto Nazionale di Fisica Nucleare, Sezione di Trieste
}
\author{
Roberto Floreanini}
\address{
Istituto Nazionale di Fisica Nucleare, Sezione di Trieste\\
Dipartimento di Fisica Teorica, Universit\`a di Trieste\\
Strada Costiera 11, 34014 Trieste, Italy
}
\maketitle
\vskip 2cm
\begin{abstract}
We study the modified dynamical evolution of
the neutral kaon system under the condition of
complete positivity.
The accuracy of the data from planned future
experiments is expected to be sufficiently precise to test 
such a hypothesis.
\end{abstract}

\vfill\eject
\narrowtext

It has been suggested that quantum gravity effects at Planck's scale
could result in loss of quantum coherence, leading to the transformation
of pure into mixed states \cite{1,2}. The neutral kaon system is a natural
subnuclear laboratory to study such phenomena \cite{3,4}. As for any decaying
system, the standard quantum time-evolution for the kaon density matrix
$\rho(t)$ is of the type 
$\partial/\partial t\rho(t)=-i\,H\,\rho(t)\,+\,i\,\rho(t)\,H^\dagger$,
where $H$ is the effective Weisskopf-Wigner Hamiltonian. This evolution
transforms pure states into pure states, although probability is not conserved:
${\rm Tr}[\rho(t)]\leq {\rm Tr}[\rho(0)]$, since $H\neq H^\dagger$.
Loss of quantum coherence shows up when the standard evolution equation
is modified as follows:
\begin{equation}
\label{1}
\frac{\partial}{\partial t}\rho(t)=\,-\,i\,H\,\rho(t)\,
+\,i\,\rho(t)\,H^\dagger+ T[\rho(t)]\quad.
\end{equation}
The non-standard (dissipative) part $T[\rho(t)]$ is a 
linear transformation 
that, in absence of Weisskopf-Wigner terms, would generate a
semigroup of maps $\tau_t$ transforming density
matrices into density matrices, preserving the positivity of their spectra,
the trace, and increasing their von Neumann entropy.
In the presence of the Weisskopf-Wigner contribution,
the trace is not preserved, but the
linear maps $\gamma_t:\rho\mapsto\rho(t)$
generated by (\ref{1}) are still positive and form a semigroup: 
$\gamma_t\circ\gamma_s=\gamma_{t+s}$, $t,s\geq0$. 
When \cite{5}-\cite{7} 
\begin{equation}
\label{2}
T[\rho(t)]=
-\frac{1}{2}\big(R\, \rho(t)+\rho(t)\, R\big)
+\sum_jA_j\, \rho(t)\, A^\dagger_j\quad,
\end{equation}
where $A_j$ and $R=\sum_jA^\dagger_jA_j$ are bounded $2\times 2$
matrices, then the time-evolution maps $\gamma_t$ are completely
positive (they form a so-called quantum dynamical semigroup and  
entropy increase is guaranteed by choosing $A_j=A^\dagger_j$ \cite{8}).

This condition is stronger than simple positivity and
has the following physical meaning \cite{9}-\cite{12}.
Let us couple the $2$-dimensional kaon system $S$ with a
$n$-dimensional system $E_n$ and let us extend the evolution maps
$\gamma_t$ to the global system $S+E_n$ in such a way that $E_n$ is
not affected: $\tilde{\gamma}_t=\gamma_t\otimes{\bf 1}_n$, ${\bf 1}_n$
being the $n\times n$ unit matrix.
Given a state $\tilde{\rho}$ of the compound system $S+E_n$, one would
like $\tilde{\gamma}_t[\tilde{\rho}]$ to be again a state,
independently of $n$ and $\tilde{\rho}$.
In general, this is not true, unless $\gamma_t$ is completely
positive.
In fact, if $\gamma_t$ is only positive, then 
states which are not of the separate form
$\tilde{\rho}=\rho_S\otimes\rho_{E_n}$ (or convex linear combinations of
these) might, for some $n$, 
develop negative eigenvalues under the action of the global
time-evolutions $\tilde{\gamma}_t$ \cite{13}.

As usual, to the time-evolution $\gamma_t$ of states of $S$ there
corresponds a dual 
time-evolution $\gamma'_t$ of the observables $X$ (bounded operators) of
$S$:
$\hbox{Tr}\big(\rho\,\gamma'_t[X]\big)=\hbox{Tr}\big(\gamma_t[\rho]X\big)$. 
The latter can be naturally extended to a linear transformation
$\tilde{\gamma}'_t:[X_{ij}]\mapsto[\gamma'_t[X_{ij}]]$ of the
observables of the compound system $S+E_n$ which are $n\times n$
matrices $[X_{ij}]$ whose entries $X_{ij}$ are observables of $S$.
Now, if $\tilde{\gamma}'_t$ transforms positive $[X_{ij}]$ into
positive operators, then $\gamma'_t$ is called $n$-positive.
If this property remains true for all $n$, then $\gamma'_t$ is called
completely positive.
In this case $\gamma'_t$ is fixed to have the general form \cite{12}
$\gamma'_t[X]=\sum_jV^\dagger_j(t)\,X\,V_j(t)$ on the observables of
$S$, for some bounded operators $V_j(t)$ such that also 
$\sum_jV^\dagger_j(t)V_j(t)$ is bounded.
By duality, the time-evolution $\gamma_t$ given by (\ref{1}) and
(\ref{2}) has the form
\begin{equation}
\label{3}
\gamma_t[\rho]=\sum_jV_j(t)\,\rho\,V^\dagger_j(t)\quad.
\end{equation}
Notice that the standard quantum mechanical evolution, arising from
(\ref{1}) when $T[\,\cdot\,]$ is absent, is as in (\ref{3}) with $j=1$ and
$V_1(t)=\exp(-i\,H\,t)$ and, therefore, is completely positive.

Completely positive maps have been used to model a large variety
of physical situations, ranging from the description of reduced
statistical systems \cite{9}-\cite{12}, \cite{14}, to the interaction of 
a mycrosystem with a macroscopic measuring apparatus \cite{15,16}, 
to a consistent description of the wave-packet reduction 
in ordinary Quantum Mechanics \cite{17}.
In the following, they will be used to 
describe the decay of the neutral kaon
system. More in general, the time evolution of 
any open quantum system can be conveniently modeled 
by a quantum dynamical semigroup.

The evolution and decay of the $K$-$\bar{K}$ system is 
conventionally modeled on a $2$-dimensional Hilbert space \cite{18}. 
In the basis of the $CP$-eigenstates, 
\begin{equation}
\label{4}
|K_1\rangle={1\over\sqrt{2}}\Big[|K\rangle+|\overline{K}\rangle\Big]\ ,\ 
|K_2\rangle={1\over\sqrt{2}}\Big[|K\rangle-|\overline{K}\rangle\Big]\ ,
\end{equation}
the generic density matrix can be expanded as 
$\rho=\rho^\mu\sigma_\mu$ in terms of the Pauli matrices $\sigma_i$,
$i=1,2,3$, and the identity $\sigma_0$.
We shall require the modified
time-evolution of the kaon system to be completely positive 
and thus generated by (\ref{1}), (\ref{2}), 
with $A_j=A^\dagger_j$ in order to ensure entropy increase.
As mentioned before, this choice is based on an effective approach to the
dynamics of the system; it has the advantage of being independent from the
details of the microscopic mechanism resposible for the loss of
quantum coherence.

On the four-vector of components $\rho^\mu$ the
dissipative term $T[\,\cdot\,]$ acts as the $4\times 4$ real, symmetric
matrix \cite{19} 
\begin{equation}
\label{6}
[T_{\mu\nu}]=-2\left(\begin{array}{cccc}
0&0&0&0\\
0&a&b&c\\
0&b&\alpha&\beta\\
0&c&\beta&\gamma
\end{array}
\right)\quad.
\end{equation}
The parameters $a$, $\alpha$ and $\gamma$ are non-negative; further,
the following inequalities must be satisfied:
\begin{eqnarray}
\nonumber
&&a\leq \alpha+\gamma\ ,\qquad 4b^2\leq \gamma^2-\big(a-\alpha\big)^2\
,\\
&&\alpha\leq a+\gamma\ ,\qquad 4c^2\leq \alpha^2-\big(a-\gamma\big)^2\
,\label{7}\\        
\nonumber
&&\gamma\leq a+\alpha\ ,\qquad 4\beta^2\leq
a^2-\big(\alpha-\gamma\big)^2\ .
\end{eqnarray}
The evolution equation (\ref{1}), with $T[\,\cdot\,]$ acting as in
(\ref{6}), generates a semigroup of completely positive maps $\gamma_t$
which could be, in principle, explicitly worked out. 
The solutions $\rho(t)\equiv\gamma_t[\rho]$ can be used to compute
certain characteristic quantities of the $K$-$\bar{K}$ system
that are directly accessible to experiments \cite{3,4}.
These are associated with the decay of neutral kaons into pion or 
semi-leptonic final states.
For instance, the decay rates into two or three pions are explicitly
given by
\begin{equation}
\label{8}
R_{2\pi}(t)={
{\rm Tr}\Big[\rho(t){\cal O}_{2\pi}\Big]\over
{\rm Tr}\Big[\rho(0){\cal O}_{2\pi}\Big]}\ ,\
R_{3\pi}(t)={
{\rm Tr}\Big[\rho(t){\cal O}_{3\pi}\Big]\over
{\rm Tr}\Big[\rho(0){\cal O}_{3\pi}\Big]}\quad,
\end{equation}
where ${\cal O}_f$ is the operator describing the final decay state $f$.
Other experimentally observed quantities are associated with the decay
of an initial $K$ state as compared to the corresponding decay of an initial
$\overline K$ state. 
These so-called asymmetries take the general form
\begin{equation}
\label{9}
A(t)={
{\rm Tr}\Big[\rho_{\bar K}(t){\cal O}_{\bar f}\Big]-
\Big[\rho_K(t){\cal O}_f\Big]\over
{\rm Tr}\Big[\rho_{\bar K}(t){\cal O}_{\bar f}\Big]+
\Big[\rho_K(t){\cal O}_f\Big]}\quad,
\end{equation}
where $\rho_{\bar K}(t)$ and $\rho_K(t)$ are the solutions of
(\ref{1}) with the initial conditions of having a pure $\bar K$
and pure $K$ at $t=0$, respectively.

On physical basis, it is plausible to assume that 
the parameters in (\ref{6}) are small, of the
order of the kaon mass squared over Planck's mass~\cite{3}.
Therefore, for the explicit computation of the quantities in (\ref{8}) and
(\ref{9}), a perturbative solution of (\ref{1}) is sufficient (the
detailed analysis can be found in \cite{19}).
A comparison of the analytic computations with the present
experimental results gives
bounds on the phenomenological parameters $a$, $b$, $c$, $\alpha$,
$\beta$ and $\gamma$ which are compatible with zero~\cite{20}.
However, more precise values for these parameters 
are expected when data from planned 
new experiments on the $K$-$\bar{K}$ system become available.
This will result into a stringent test of the inequalities (\ref{7}), and
hence of complete positivity as a viable physical condition on possible
modifications of the neutral kaon dynamics. To our knowledge, this
is the first time that complete positivity can be put to
test to such a high accuracy in the study of a subnuclear system. 

The new experiments on the $K$-$\bar K$ system will also be able to
detect physical inconsistencies that might arise when the map
$T[\,\cdot\,]$ in (\ref{1}) is taken to be simply positive and
not completely positive.
An example is given by maps $T[\rho(t)]$ in (\ref{1}) 
of the form (\ref{6}) with $a=b=c=0$, $\alpha\neq\gamma$,
$\beta\neq 0$ and $\alpha\gamma\geq\beta^2$.
Such a modification of the standard quantum mechanical treatment of
the neutral kaon dynamics has been proposed and discussed in \cite{3,4}.
In this case, (\ref{1}) generates a semigroup of simply positive
dynamical maps $\gamma_t$.
In fact, due to the inequalities (\ref{7}), complete positivity
would require $\alpha=\gamma$ and $\beta=0$, thus providing a
trivial $T[\,\cdot\,]$.~\footnote{
Notice that the inequalities~(\ref{7}) put stringent bounds on
possible hierarchies between the non-standard parameters.
In particular, the possibility of recovering from the
completely positive approach the simply positive one discussed 
in~\cite{1}--\cite{4} as an effective description by making
$a$, $b$, $c$ very small with respect to
$\alpha$, $\beta$ and $\gamma$ seems to be ruled out~\cite{21},~\cite{20}.}
\vfill\break

In order to study the properties of such simply positive maps
$\gamma_t$,
let us first consider the time-evolution 
$\tau_t[\rho]=\exp(t\,T[\,\cdot\,])[\rho]$
generated by  the non-standard part $T[\,\cdot\,]$.
Its action on density matrices 
$\rho=\rho^\mu\sigma_\mu$ written as
four-vectors of components $\rho^\mu$
can be explicitly worked out starting from (\ref{6}) with $a=b=c=0$;
it is given by the $4\times 4$ matrix
\begin{equation}
\label{10}
[\tau_{\mu\nu}]=\left(\begin{array}{cccc}
1&0&0&0\\
0&1&0&0\\
0&0&A(t)&B(t)\\
0&0&B(t)&C(t)
\end{array}
\right)\quad,
\end{equation}
where 
\begin{eqnarray}
\nonumber
&&
A(t)=\frac{1}{\lambda_+-\lambda_-}\left[
(\lambda_++2\alpha)\,e^{\lambda_-\,t}-
(\lambda_-+2\alpha)\,e^{\lambda_+\,t}\right]\ ,\\
\label{10b}
&&
B(t)=\frac{2\beta}{\lambda_+-\lambda_-}
\left(e^{\lambda_-\,t}-e^{\lambda_+\,t}\right)\ ,\\
\nonumber
&&
C(t)=\frac{1}{\lambda_+-\lambda_-}\left[
(\lambda_++2\alpha)\,e^{\lambda_+\,t}-
(\lambda_-+2\alpha)\,e^{\lambda_-\,t}\right]\ ,
\end{eqnarray}
and $\lambda_\pm=-(\alpha+\gamma)\pm\sqrt{(\alpha-\gamma)^2+4\beta^2}$
are both negative due to the positivity condition $\alpha\gamma\geq\beta^2$.
If $\tau_t$ were completely positive, we should be able to write
its action as in (\ref{3}); this is possible only when
$\lambda_+=\lambda_-$ or equivalently when $\alpha=\gamma$, $\beta=0$.

The map $\tau_t$ is an example of a positive linear transformation
which is not $2$-positive.
Indeed, as previously discussed, let us extend $\tau_t$ to 
$\tilde{\tau}_t=\tau_t\otimes{\bf 1}_2$ acting on the two-dimensional
kaon system coupled to another arbitrary two-dimensional system.
Then, $\tilde{\tau}_t$ transforms the (entangled) state 
\begin{equation}
\label{11}
\rho_S=\frac{1}{4}\Big(
\sigma_0\otimes\sigma_0-\sum_{i=1}^3\sigma_i\otimes\sigma_i\Big)\ ,
\end{equation}
into
\begin{eqnarray}
\nonumber
&&\tilde{\tau}_t[\rho_S]=\frac{1}{4}\Big(
\sigma_0\otimes\sigma_0-\sigma_1\otimes\sigma_1
-A(t)\,\sigma_2\otimes\sigma_2\\
\label{12}
&&\ \ -C(t)\,\sigma_3\otimes\sigma_3-B(t)\big(\sigma_2\otimes\sigma_3
+\,\sigma_3\otimes\sigma_2\big)\Big)\ .
\end{eqnarray}
This operator is no more positive as can be seen by computing its mean
value on the states 
\begin{equation}
\label{13}
\begin{array}{l}
|u\rangle=\left(\begin{array}{c}
1\\
0
\end{array}\right)\otimes
\left(\begin{array}{c}
1\\
0
\end{array}\right)
+
\left(\begin{array}{c}
0\\
1
\end{array}\right)\otimes
\left(\begin{array}{c}
0\\
1
\end{array}\right)\\
\\
|v\rangle=
\left(\begin{array}{c}
1\\
0
\end{array}\right)\otimes
\left(\begin{array}{c}
0\\
1
\end{array}\right)
+
\left(\begin{array}{c}
0\\
1
\end{array}\right)\otimes
\left(\begin{array}{c}
1\\
0
\end{array}\right)\ .
\end{array}
\end{equation}
Explicitly, one finds 
\begin{equation}
\label{14}
\langle u|\tilde{\tau}_t[\rho_S]| u\rangle=
-\langle v|\tilde{\tau}_t[\rho_S]| v\rangle=
\frac{1}{2}\big(A(t)-C(t)\big)\ .
\end{equation}
This quantity never vanishes for $t>0$ and has a definite sign
depending on the relative magnitude of $\alpha$ and $\gamma$; therefore, 
if $A(t)-C(t)$ is positive in one case, it is negative
in the other.
This means that $\tilde{\tau}_t[\rho_S]$ develops negative eigenvalues.
If one adds to the generator $T[\rho]$ the Weisskopf-Wigner part as in
(\ref{1}), this result is not altered as we shall see later.

At first sight, this conclusion might seem of little physical
relevance.
Indeed, the coupling of the subsystem of interest to 
an  abstract $n$-level system is regarded as too artificial by
those who consider the condition of complete positivity of the reduced
dynamics as a mere technical request \cite{22}.
The point is that in our case, the additional two-dimensional system
can be taken to be another kaon system.
Precisely this physical situation is commonly encountered in the
so-called $\phi$-factories.  
In these experimental setups, a $\phi$ meson decays into an
entangled antisymmetric state of two neutral kaons that in our
formalism can be written as \cite{4}
\begin{equation}
\label{15}
|\Psi_S\rangle=\frac{1}{\sqrt{2}}\big(|K_1\rangle\otimes|K_2\rangle-
|K_2\rangle\otimes|K_1\rangle\big)\ .
\end{equation}
As a density matrix $|\Psi_S\rangle\langle\Psi_S|$ 
this state exactly corresponds to the
``singlet'' $\rho_S$ in (\ref{11}).

The global time-evolution of states like (\ref{15}) is obtained by the
tensor product $\Omega_t=\gamma_t\otimes\gamma_t$ of the
time-evolutions of the single kaons.
In the standard approach, with purely Weisskopf-Wigner dynamics,
positive operators remain positive and negative eigenvalues are
never generated.
Indeed, this is guaranteed by any completely positive time-evolution 
$\gamma_t$ as comes out from (\ref{3}):
\begin{equation}
\Omega_t[\rho]=\sum_{ij}\big[V_i(t)\otimes V_j(t)\big]\,\rho\,
\big[V^\dagger_i(t)\otimes V^\dagger_j(t)\big]\ .
\label{16}
\end{equation}
On the contrary, the positivity of the evolving states need not be 
preserved by modified dynamics $\gamma_t$ that are only simply positive. 
Again, the single kaon time-evolution given 
by (\ref{10}) provides us with an example of this fact.
The action of ${\cal T}_t=\tau_t\otimes\tau_t$ on 
$|\Psi_S\rangle\langle\Psi_S|=\rho_S$ gives
\begin{eqnarray}
\nonumber
&&{\cal T}_t\big[\rho_S\big]=\frac{1}{4}\Big[
\sigma_0\otimes\sigma_0-\sigma_1\otimes\sigma_1\\
\nonumber
&&-\big(A^2(t)+B^2(t)\big)\sigma_2\otimes\sigma_2-\big(B^2(t)+C^2(t)\big)
\sigma_3\otimes\sigma_3
\\
\label{17}
&&-B(t)\big(A(t)+C(t)\big)
\big(\sigma_2\otimes\sigma_3+\sigma_3\otimes\sigma_2\big)\Big]\ ,
\end{eqnarray}
and, therefore, the presence of negative eigenvalues can be
ascertained as before by computing the mean values
on the vectors (\ref{13}):
\begin{eqnarray}
\nonumber
\langle u|{\cal T}_t\big[\rho_S\big]| u\rangle&=&
-\langle v|{\cal T}_t\big[\rho_S\big]|
v\rangle\\
\label{18}
&=&\frac{1}{2}\big(A^2(t)-C^2(t)\big)\neq0\ .
\end{eqnarray}
We now show that this pathology cannot be cured by considering also
the Weisskopf-Wigner contribution to the time-evolution.
Let us call ${\cal W}_t$ the full time-evolution $\Omega_t$
when the non-standard dissipative part is absent.
It corresponds to the ordinary quantum mechanical evolution on the
compound two-kaon system.
Then, $\Omega_t$ can be obtained from ${\cal W}_t$ and ${\cal T}_t$
through the Trotter product formula
\begin{equation}
\label{19}
\Omega_t[\rho]=\lim_{n\to\infty}
\left({\cal W}_{t/n}\circ{\cal T}_{t/n}\right)^n[\rho]\ .
\end{equation}
As already noticed the Weisskopf-Wigner evolution
${\cal W}_t$ preserves the positivity of operators.
From (\ref{10}) it can be checked that ${\cal T}_t$ is trace
and hermiticity preserving.
Further, as proved above, the operator
${\cal T}_{t/n}\big[\rho_S\big]$
has negative eigenvalues and therefore can be written as the sum
of non-trivial positive and negative parts which are not altered
by the action of ${\cal W}_{t/n}$.
Therefore, we can write:
\begin{equation}
\label{20}
\left({\cal W}_{t/n}\circ{\cal T}_{t/n}\right)[\rho_S]=\rho_++\rho_-\ ,
\end{equation}
where $\rho_+$ is a positive and $\rho_-$ a negative operator.
The iteration of this action cannot destroy the presence of a
negative part.
In fact, since ${\cal T}_t$ is trace preserving one has
\begin{eqnarray}
\nonumber
\hbox{Tr}(\rho_-)&=&\hbox{Tr}\left({\cal T}_{t/n}[\rho_-]\right)\\
\label{20b}
&=&
\hbox{Tr}\left(\left[{\cal T}_{t/n}[\rho_-]\right]_+\right)+
\hbox{Tr}\left(\left[{\cal T}_{t/n}[\rho_-]\right]_-\right)\ .
\end{eqnarray}
Therefore, since 
$\hbox{Tr}\left(\left[{\cal T}_{t/n}[\rho_-]\right]_+\right)\geq 0$,
while the trace of any negative operator is obviously $\leq 0$,
it turns out that
\begin{equation}
\label{20c}
\left|\hbox{Tr}\left(\left[{\cal T}_{t/n}[\rho_-]\right]_-\right)\right|
\geq \left|\hbox{Tr}(\rho_-)\right|\ .
\end{equation}
Further, notice that
\begin{eqnarray}
\nonumber
\frac{\partial}{\partial s}\hbox{Tr}&&\left( 
{\cal W}_s\left[\left[{\cal T}_{t/n}[\rho_-]\right]_-\right]\right)\\
\label{20d}
&&=-
\hbox{Tr}\left(
\big(\Gamma\otimes{\bf 1}_2+{\bf 1}_2\otimes\Gamma\big)
{\cal W}_s\left[\left[{\cal T}_{t/n}[\rho_-]\right]_-\right]\right)\\
\nonumber
&&\leq-2\lambda\,\hbox{Tr}\left(
{\cal W}_s\left[\left[{\cal T}_{t/n}[\rho_-]\right]_-\right]\right)\ ,
\end{eqnarray}
where $\Gamma$ is the width matrix of the Weisskopf-Wigner
hamiltonian and $\lambda$ its largest eigenvalue.
Integrating this inequality  with respect to $s$ from $0$ to $t/n$,
one obtains
\begin{equation}
\label{20e}
\hbox{Tr}\left( 
{\cal W}_{t/n}\left[\left[{\cal T}_{t/n}[\rho_-]\right]_-\right]\right)\leq
e^{-2\lambda\,t/n}
\, \hbox{Tr}\left(\left[{\cal T}_{t/n}[\rho_-]\right]_-\right)\ .
\end{equation}
Putting together (\ref{20c}) and (\ref{20d}), one gets
\begin{equation}
\label{20f}
\left|\hbox{Tr}\left(\left[\left(
{\cal W}_{t/n}\circ{\cal T}_{t/n}\right)[\rho_-]\right]_-\right)\right|\geq\,
e^{-2\lambda\,t/n}\left|\hbox{Tr}(\rho_-)\right|\ .
\end{equation}
In deriving this formula we have used the fact
that $\left[{\cal W}_t[\rho]\right]_-={\cal W}_t[\rho_-]$, for any $\rho$;
this follows from the invertibility and the positivity preserving property of
$\cal W$ and from the uniqueness of the decomposition of self-adjoint 
operators into positive and negative parts.

Iterating $n-1$ times the above estimates, one finally obtains 
\begin{equation}
\label{21}
\left|\hbox{Tr}\left(\left[\left({\cal W}_{t/n}\circ{\cal T}_{t/n}\right)^n
[\rho_S]\right]_-\right)\right|\geq\,
e^{-2\lambda\,t}\, \left|\hbox{Tr}(\rho_-)\right|>0\ .
\end{equation}
Therefore, the  negative part in 
$\left({\cal W}_{t/n}\circ{\cal T}_{t/n}\right)^n
[\rho_S]$
will survive the limit $n\to\infty$; hence 
$\Omega_t[\rho_S]$ will always have
negative eigenvalues.
In other words, the full time-evolution $\Omega_t$ on
the correlated ``singlet'' state of the two kaons will always generate
negative probabilities.

In conclusion, we have discussed how the condition of
complete positivity on the time-evolution might be tested in actual
experiments involving neutral kaons.
On one hand, this can be done
by verifying certain inequalities that must be fulfilled
by the parameters of any completely
positive modification of the standard phenomenological
Weisskopf-Wigner description of a single kaon system.
On the other hand, by studying time-correlations of entangled two kaon
systems one can distinguish between completely positive {\it vs}
simply positive modified dynamics due to the emergence, in the
latter case, of negative probabilities.

\end{document}